\documentclass[12pt]{article}
\usepackage{graphicx}
\usepackage{color}
\usepackage{amsmath,slashed, amssymb}
\usepackage{fancyhdr}
\usepackage{hyperref}
\usepackage[titletoc]{appendix}
\numberwithin{equation}{section} 

\def\fnote#1#2{\begingroup\def\thefootnote{#1}\footnote{#2}\addtocounter
{footnote}{-1}\endgroup}

\def\beq{\begin{eqnarray}}
\def\eeq{\end{eqnarray}}
\def\bea{\begin{eqnarray*}}
\def\eea{\end{eqnarray*}}

\def\mh{m^{(1)}_{hol}}
\def\mmh{m^{(2)}_{hol}}
\def\mih{m^{(i)}_{hol}}

\def\NPB#1#2#3{Nucl. Phys. {\bf B#1}, #3 (#2)}
\def\PLB#1#2#3{Phys. Lett. {\bf B#1}, #3 (#2)}
\def\PLBold#1#2#3{Phys. Lett. {\bf #1B}, #3 (#2)}
\def\PRP#1#2#3{Phys. Rep. {\bf #1}, #3 (#2)}
\def\PRD#1#2#3{Phys. Rev. {\bf D#1}, #3 (#2)}
\def\PRold#1#2#3{Phys. Rev. {\bf #1} (#2) #3}
\def\PRL#1#2#3{Phys. Rev. Lett. {\bf #1}, #3 (#2)}
\def\PREP#1#2#3{Phys. Rep. {\bf #1} #3, (#2)}
\def\ZPC#1#2#3{Z. Phys. C {\bf #1}, #3 (#2)}

\def\centeron#1#2{{\setbox0=\hbox{#1}\setbox1=\hbox{#2}\ifdim
\wd1\rangle\wd0\kern.5\wd1\kern-.5\wd0\fi
\copy0\kern-.5\wd0\kern-.5\wd1\copy1\ifdim\wd0\rangle\wd1
\kern.5\wd0\kern-.5\wd1\fi}}
\def\ltap{\;\centeron{\raise.35ex\hbox{$\langle$}}{\lower.65ex\hbox{$\sim$}}\;}
\def\gtap{\;\centeron{\raise.35ex\hbox{$\rangle$}}{\lower.65ex\hbox{$\sim$}}\;}
\def\gsim{\mathrel{\gtap}}
\def\lsim{\mathrel{\ltap}}

\def\doublespaced{\baselineskip=\normalbaselineskip\multiply
    \baselineskip by 200\divide\baselineskip by 100}
\def\singleandhalfspaced{\baselineskip=\normalbaselineskip\multiply
    \baselineskip by 150\divide\baselineskip by 100}
\def\singleandabitspaced{\baselineskip=\normalbaselineskip\multiply
    \baselineskip by 120\divide\baselineskip by 100}
\def\singleandthirdspaced{\baselineskip=\normalbaselineskip\multiply
    \baselineskip by 130\divide\baselineskip by 100}
\def\singlespaced{\baselineskip=\normalbaselineskip}

\newcommand{\newc}{\newcommand}
\newc{\qbar}{{\overline q}}
\newc{\Kahler}{Kahler }
\newc{\deltaGS}{\delta_{\rm GS}}

\def\lrf#1#2{ \left(\frac{#1}{#2}\right)}
\def\lrfp#1#2#3{ \left(\frac{#1}{#2} \right)^{#3}}
\newcommand{\la}{\left\langle}
\newcommand{\ra}{\right\rangle}
\newcommand{\lp}{\lambda_{p}}
\newcommand{\Fg}{\mathcal{F}_G}
\newcommand{\Fgp}{\mathcal{F}_{G,\,p}}
\newcommand{\Fgpd}{\overline{\mathcal{F}}_{G,\,p}}
\newcommand{\Dgp}{D_{G,\,p}}
\newcommand{\Dg}{D_G}
\newcommand{\pp}{\phi_{p}}
\newcommand{\Fp}{F_{p}}

\begin{document}
\begin{titlepage}
\begin{flushright}
{\large SCIPP 18/02\\
\large ACFI-T17-08 \\}
\end{flushright}

\vskip 1.2cm

\begin{center}

{\LARGE\bf Comments on the Transplanckian Censorship Conjecture}

\vskip 1.4cm

{\large Michael Dine and Yan Yu}
\\
\vskip 0.4cm
{\it $^{(a)}$Santa Cruz Institute for Particle Physics and
\\ Department of Physics, University of California at Santa Cruz \\
     Santa Cruz CA 95064  } \\
     ~\\
\vspace{0.3cm}

\end{center}

\vskip 4pt


\begin{abstract}
We consider some aspects of the {\it Transplanckian Censorship Conjecture} (TCC), which states that for theories of quantum gravity there is a limit on the lifetime of dS or quintessence states
not too different than the current Hubble horizon.  If one accepts the {\it de Sitter Swampland conjecture}, then the former are ruled out.  We consider some aspects of tunneling to an isolated ground state in the presence of time-varying fields, in 
quantum mechanics and quantum field theory in the absence of gravitation, and note that lifetimes are typically enormous; in fact, there is often a finite probability for the system to remain eternally in its original
state.  With gravity in a universe with superluminal expansion, while
the field evolution may be slowed, Planck scale fluctuations would seem likely to grow to
superhorizon size long before the universe decays.  We argue that the TCC, if it is correct, requires that superluminal expansion occur only for a brief period in the history of the universe, and will be followed by
a $p=\rho$ phase.
\end{abstract}

\end{titlepage}

\section{Introduction:  The de Sitter Swampland Conjecture and Quintessence}

It has proven challenging to construct de Sitter solutions of string theory, and this has motivated the {\it de Sitter Swampland Conjecture}\cite{ovswampland}.  While the two present authors argued in \cite{desitterchallenges} 
that this is a question which is intrinsically inaccessible to current controlled approaches to string theory, in this paper we will generally assume that the conjecture is correct, and review and further explore some of its consequences.  As discussed in \cite{ovswampland}, one of the most immediate is that the observed dark energy must be a form of quintessence.   Quintessence requires equation of state parameter $w < -1/3$.  Observations require $w$ close to $-1$.  Whether such configurations arise in string theory -- with sensible scales of energy, mass scales for the Higgs, and so on -- is an interesting (and extremely challenging) question, which we will not explore here.  We will confine ourselves to a narrower question.  Assuming that one has a landscape in string theory with quintessence like configurations as well as negative cosmological constant stationary points, then these states are unstable to decay to AdS spaces (more precisely to states of negative cosmological constant).  Ref. \cite{tcc} insisted on a criterion for sensible states\footnote{In the context of inflation, possible limits on the number of e-folds have been discussed by others, e.g. \cite{efoldlimits,padilla}. On the other hand, 
the authors of \cite{burgess} present arguments that such transplanckian fluctuations might be consistent with successes of inflation}.  They noted that in these quintessence states, one has superluminal expansion.  If the state lives long enough, Planck scale fluctuations will redshift until they become larger than the horizon.  The authors of \cite{tcc} argued that this is not sensible, and that there should
be a limit on the lifetime, $\Gamma$, of such states:
\beq
\Gamma < H \log(H).
\eeq

While not committing ourselves to a view on this basic question of principle, we will ask:  is this plausible?  What are the lifetimes of such quintessence states likely to be?  This will require that we consider
some aspects of tunneling, from a state which is evolving in time towards zero energy, to a lower energy state.  We will first consider this in quantum mechanics, and then in quantum field theory without gravity, and finally in a generally relativistic theory.  In the first two cases, we will see that the lifetimes
can be quite long.  In fact, typically, there is a finite probability that the system never makes a transition to the lower state at all.  These systems are not amenable
to conventional WKB/Euclidean path integral approaches, but it is not difficult to make rough estimates of the tunneling rates working with Minkowski signature.

Including general relativity, and more generally in a would-be quantum
theory of gravity, we are on less certain ground.
In string theory, in the absence of supersymmetry, one might expect that states with potentials which fall to zero for large values of scalar (``pseudomoduli") fields are typical.  Such potentials would go to zero at least exponentially rapidly in various regions of field space.   We will generally model the tunneling problem by considering potentials which, in one direction direction, labeled by a field
$\phi$, are pure exponentials, $V(\phi) = A e^{-\lambda \phi}$.\footnote{In terms of fields with canonical kinetic terms, potentials might well be expected to tend to zero far more rapidly.}  There is another direction,
$\chi$, such that at some point, there is a local minimum with negative
c.c. for both $\lambda$ and $\chi$.  As we will review, in the first regime, unless the
coefficient in the exponential satisfies a certain bound, the potential quickly becomes negligible and one has a universe with $p=\rho$, i.e. $w=1$.   The TCC does not constrain these systems.  For sufficiently small $\lambda$, the system may exhibit quintessence.  We will focus our considerations on such states.  We will argue that tunneling is highly
suppressed, as in the non-gravitational case.

The rest of this paper is organized as follows.  In section \ref{swamplandconjecture}, we review the de Sitter swampland conjecture and the Trans Planckian censorship conjecture.  We note the conditions on exponential
potentials to obtain $w < -1/3$.  In section \ref{quintessencelandscape}, we discuss what a landscape might look like which possesses quintessence states and AdS minima.  We provide some model potentials on this landscape, which will guide our tunneling computations.  In section \ref{quantummechanics}, we consider tunneling in quantum mechanical models with small numbers of degrees of freedom, between states
with one degree of freedom ``rolling down a hill", while in another direction, the potential exhibits a minimum.   Then we treat the analogous problem in field theory, first in the absence of general relativity
in section \ref{fieldtheory}, and then coupling to general relativity in section \ref{withgravity}.   Finally we
turn to conjectures about how such tunneling might look in a landscape.  We are particularly concerned with how the requirement of slow variation of the potential, with resulting slow motion of the field,
might allow different behaviors than in our quantum mechanics and field theory examples.   We conclude in section \ref{conclusions}, that the TCC is likely incompatible with the pictures which have been put forth for a landscape.

\section{The de Sitter Swampland And Trans Planckian Censorship Conjectures}
\label{swamplandconjecture}

The vacuum states we can claim to understand in string theory generally possess a high degree of supersymmetry.  States without supersymmetry, especially de Sitter space (or flat space) are hard to access by weak coupling methods\cite{dineseibergproblem}.  Indeed, as stressed
in \cite{desitterchallenges}, typical non-supersymmetric states
exhibit runaway to singular space-times and strong coupling, and
one cannot claim to understand these in any systematic way.  Such states might be candidates for quintessence.  That said,
the work of Bousso and Polchinski\cite{boussopolchinski} and KKLT\cite{kklt} suggests the possible existence of a landscape of states, with a discretuum of positive and negative cosmological constants.  The existence
of these states can hardly be viewed as rigorously established.  Ooguri and Vafa\cite{ovswampland} conjectured, based on the difficulty of finding dS stationary points of
effective potentials of systems with branes and fluxes\cite{andriot}, that de Sitter vacua, stable or unstable, may not exist, and that the presently
observed dark energy is a form of quintessence.  In \cite{desitterchallenges}, it was demonstrated that
there are fundamental obstacles to weak coupling searches, and argued that these don't provide an argument, one way or the other, about the existence
of metastable de Sitter space in string theory/quantum gravity.  The work of that reference focused heavily on the fact that such would-be states are metastable, and in the past or future, the space-time becomes singular at the classical level.

In \cite{tcc}, Bedroya and Vafa set forth an additional conjecture.   They argued that Planck scale fluctuations should not become classical.  So for any
would-be state, the lifetime, $T$, should satisfy
\beq
T< H_f^{-1} \log H_f,
\eeq 
 where $H_f$ is the Hubble parameter at the moment of decay.  This decay might represent a point in time where superluminal expansion ends.  We will not focus on this possibility, though for $w$ close to $-1$,
 such a situation might well be tuned.  It might also represent a moment of vacuum decay, in the sense of \cite{cdl}, which will be the focus of our current discussion.
 
 If a landscape picture holds, any quintessence vacuum will be surrounded by classically stable, negative cosmological constant stationary points.  As the
 quintessence field roles in its potential, it can decay to one of these stable minima, but we would expect that the decay amplitude would rapidly get smaller
 as the field $\phi$ rolls down its hill.  This is a slightly unconventional tunneling problem, and we will consider it first in a quantum mechanics system
 with two degrees of freedom, and then in field theory without general relativity, before attacking the actual problem of interest.  In both of these cases, we will find significant suppression
 of the decay amplitude.  Turning to the gravitational case with
 quintessence, we lack an explicit, controlled string model.  Still, it would seem that if such systems exist, lifetimes, for large values of teh quintessence field/small values of the energy density are likely to be very long, violating the conjecture.  So there would
 seem to be a tension between the TCC and a landscape picture.

\section{Quintessence in a Landscape}
\label{quintessencelandscape}

First, we recall some basic facts about quintessence.  The equation of state
$p = w \rho$
leads to evolution of the scale factor, according to:
\beq
a(t) = a(t_0) ({t \over t_0}) ^{2 \over 3(1+ w)}.
\eeq
For $w=1$, a free massive field,  $$a(t) \propto t^{1/3}.$$  More generally, this result holds when, in some era, one can neglect the potenital.
$w \le -1/3$ leads to quintessence; $a(t)$ grows faster than $t \sim {1 \over H}$.

We will focus on $w < -1/3$, considering a field $\phi$ with a canonical kinetic term and an exponential potential,
\beq
V(\phi) = \Lambda^4 e^{-\lambda \phi}
\eeq
in units with reduced Planck mass, $\tilde M_p  = 1$.
For $\lambda < \sqrt{6}$
\beq
w = -1+ { \lambda^2 \over 3}
\eeq
with $\lambda < \sqrt{2}$ necessary for quintessence\cite{quintessencereview}.  One can quickly check this formula
for small $\lambda$, noting that in this case the second derivative
in the equation for $\phi$, 
\beq
\ddot \phi + {3  H} \dot \phi + V^\prime(\phi)
\eeq
can be neglected. So $T_{00}$ and $T_{ij}$ can be computed
simply for the exponential potential.


So now the interesting question is:  suppose one has a string model with such a potential and that this accounts for the observed dark energy.  In a landscape context, we expect that there are states nearby with
negative cosmological constant.  We can model this by considering two fields, $\phi$ and $\chi$, with potential such that $\chi = 0$, $\phi>0$ corresponds to the quintessence state, and $\chi = \mu,~\phi= 0$ corresponds to an AdS minimum.  The TCC raises the question: what is the lifetime of the quintessence state? 

We should note that, for  the exponential potential, assuming $V$ is as small as the dark energy today, and that $\Lambda$ is not extremely small, $\phi$ is large, so it is likely that there are many
light states.  For the question of vacuum decay, this feature would seem likely to further suppress the decay rate.  How this might effect the criteria for quintessence we won't explore, but it might increase the
tuning required to obtain the required brief period of superluminal expansion.








\section{Tunneling from Quintessence-Like states in Quantum Mechanics}
\label{quantummechanics}

We first consider a quantum mechanical problem, with two degrees of freedom, $\chi$ and $\phi$, which exhibits tunneling from
a time-dependent configuration of the coordinates.  In particular, classically there is a lowest energy configuration for some value of $(\chi,\phi) = (\mu,0)$ and a higher energy configuration, where $\chi=0$ and $\phi$ is not uniquely fixed, but instead the potential falls to zero
for large $\phi$ and $\chi= 0$:
\beq
    V(\chi,\phi) = \lambda \phi^2(\chi)^2 + \Gamma(\chi ^2-\mu^2)^2 + \delta\frac{(\chi -\mu)^2}{\phi^2 + \mu^2}
    \label{chiphipotential}
\eeq
This has a global minimum at $\chi= \mu$, $\phi = 0,~V=0$.  At $\chi=0$, it has runaway behavior for $\phi$.  
 %

The false ``minimum" has $\chi=0$, $\phi$ rolling, with
\beq
V(\phi)=\delta {\mu^2 \over \phi^2 + \mu^2}
\label{phichipotential}
\eeq
(Note we are assuming here that $\delta < \mu^4$; other parameters have been chosen for simplicity; small changes will not alter the behavior
of the potential).

This model suggests focusing on a single degree of freedom, $x$, with:
\beq
V(x)= -V_0~~~x <x_0~~~~~V(x) = {\delta \over x^2} ~x> > x_0
\label{modelpotential}
\eeq
We are interested in tunneling from a configuration described by a wave packet centered at $x>x_0$, and evolving with time.
If the wave packet is
Gaussian, and sufficiently narrow, there will be a huge suppression of the wave function in the region $x < x_0$.  The question is:  what is the natural value for this width, and how does the width
grow with time.

It is worth recalling some facts familiar from elementary quantum mechanics.
\begin{enumerate}
\item  Wave packet evolution for a free particle:    consider a system described at $t=0$ by a Gaussian wave packet with width $\Delta x$,
\beq
\psi(x,0) = e^{i k_0 x}e^{-{(x-x_{cl})^2 \over (\Delta x)^2}} .   
\eeq
A standard approach to this problem is to Fourier transform, use the known behavior of plane waves, and Fourier transform back.  In this case,
one obtains:
\beq
\psi(x,t) = e^{i k_0 x - i {k_0^2 \over 2m} t} e^{-{(x-x_{cl}(t))^2 \over (\Delta x)^2  + i {t \over m}} }  ~~x_{cl}(t) = x_0 + {k_0 \over m} t  
\eeq
So the width of the wave packet grows with time according to:
\beq
(\Delta x(t))^4 = (\Delta x)^4 + {t^2 \over m^2}     
\label{xwidth}
\eeq
This is what one expects from a simple-minded semiclassical argument.
With the passage of time, the width of the packet grows as $(\Delta v) t={\Delta k \over m} t$, giving
\beq
(\Delta x(t))^2 = (\Delta x)^2 + {(\Delta k)^2 t^2 \over m^2}
\eeq
or
\beq
(\Delta x(t)^4) = (\Delta x)^4 + 2 {t^2 \over m^2 } + {\cal O} (t^4).    
\eeq
From equation \ref{xwidth}, we have, at large times,
\beq
    \Delta x(t) = \sqrt{t \over m},
\eeq
independent of the initial width of the packet.  The width grows much more slowly than the packet moves, i.e.
\beq
{d \Delta x(t) \over dt} \ll v.
\eeq{}
\item  Wave packet evolution for a harmonic oscillator:  here, the standard textbook result is that the center of the wave packet evolves classically, and the wave packet does not spread in time.  We can see this directly
in coordinate space. With
\beq
\psi(x,t) = e^{ - i {k_0^2 \over 2m} t} e^{-{(x-x(t))^2 \over (\Delta x)^2 } }  ~~x(t) = A \cos(\omega t).
\eeq
In the Schrodinger equation, we can compare the term $-{1 \over 2m}{\partial^2 \psi \over \partial x^2}$ with the $K x^2$ term.  This fixes $(\Delta x)^2 = {1 \over \omega^2}$.  The wave packet does not spread.
\item  Our problem is presumably somewhere in between, behaving nearly
like a free particle, with the wave packet spreading, perhaps somewhat more slowly than that for a free particle, particularly in the region where the potential is growing.
\beq
{\cal A}\propto e^{-{x(t)^2 \over \Delta x^2}}.
\eeq
\end{enumerate}
If this is the case, for large times, we have a huge suppression of the tunneling amplitude,
Rather than a rate of decay per unit time, the rate falls exponentially to zero at large times; one has simply a finite probability to remain in the rolling condition forever.

.

\section{Tunneling from Quintessence-Like states in Field Theory (Without Gravity)}
\label{fieldtheory}

For the analogous problem in field theory, we consider two fields, $\chi$ and $\phi$, with the potential of equation \ref{chiphipotential}
\beq
    V(\chi,\phi) = \lambda \phi^2(\chi)^2 + \Gamma(\chi ^2-\mu^2)^2 + \delta\frac{(\chi -\mu)^2}{\phi^2 + \mu^2}
    \label{chiphipotentialfieldtheory}
\eeq
Again, we have an isolated vacuum at $\phi = 0, \chi=\mu$, and a quintessence-like configuration at $\chi=0$. For
$\phi >  \mu$, the $\chi$ potential gets steeper and steeper for larger $\phi$.  Since our interest is in estimating the decay
from the region of large $\phi$, it makes sense to model the system integrating out $\chi$ and writing
\beq
V_{model}(\phi) = -V_0 ~~\phi < \mu;~~V(\phi) ={\delta \over \phi^2} ~~\phi> \mu.
\eeq
We want to investigate, again, the decay of the quintessence-like state to the isolated vacuum (we can smooth out $V_{model}$ around
$\phi = \mu$, if desired).

In terms of the model potential, we can describe the initial bubble corresponding to decay of the system, if we treat $\phi(t)$ as fixed, and take
an even more drastic simplification of the potential:
\beq
V_{simplified}(\phi) = -V_0 ~~\phi < \mu;~~V(\phi) =0 ~~\phi> \mu.
\eeq
Now we might expect the critical bubble, on dimensional or simple scaling grounds, to have radius:
\beq
R^2 = \phi(t)^2/V_0.
\eeq
Specifically, the kinetic energy term would be of order $\phi(t)^2 R$, while the potential energy term would be of order $R^3 V_0$; the balance
determines $R$.
Since the field, $\phi$, is free mostly everywhere, we might expect it to have a Gaussian wave functional,
\beq
\Psi(\phi) = e^{-\int d^3 x \phi(x) (\nabla^2) \phi(x)}
\eeq
and correspondingly the amplitude to find such a bubble would behave as
\beq
{\cal A} \sim e^{-\phi^2 R^2} \sim e^{-\phi^4/V_0}.
\label{tunnelingamplitude}
\eeq
This is, of course, extremely suppressed at large $\phi$.

We can obtain the estimate of equation \ref{tunnelingamplitude} by a WKB analysis as in the thin wall case.  If we think of a ``standard bubble" with size of order $R$ and variations of $\phi$ on scales of order $R$, the
lagrangian for $R$ is now:
\beq
L = \phi_0^2 R \dot R^2 + V_0 r^2 - \phi_0^2 R.
\eeq
The critical point in the potential is then
\beq
R^2 = V_0^{-1} \phi_0^2
\eeq
so the WKB estimate yields
\beq
B \sim \int dR R^{1/2} \sqrt{V_0 R^3}
\sim {\phi_0^4 \over V_0}
\eeq
as above.

\section{Including Gravity and Checking the TCC Citerion}
\label{withgravity}

Without gravity, for large $\phi_0$, we have seen that the amplitude for bounce production
is enormously suppressed at large $\phi$.  Here we ask the extent to which gravity might
qualitatively alter these results.

\subsection{The Size of Gravitational Corrections}

It is worth considering, first, the size of such corrections in the famliar case of thin wall tunneling\cite{cdl}, for small gravitational coupling, $G_N$, and with energy splitting $\epsilon$ and critical bubble radius $R$.  For the case of decay from flat space to de Sitter space, the energy of the bubble is of order $\epsilon R^3$, and the gravitational field is of order $G_N \epsilon R^2$.
So the corrections to the action for $R$ are of order $G_N \epsilon^2 R^5$,
consistent with equation 3.19 of ref. \cite{colemandecay}.  These effects
grow as $\epsilon$ becomes smaller, for fixed $G_N$, and can be quite dramatic, as stressed by Coleman and Deluccia.  For the field theory systems described in the last section, this can also be true for very large $\phi(t)$, but there is a period where these corrections are under control and small.  In this period the tunneling amplitude is extremely tiny.

It is in the presence of gravity that we can can actually
discuss quintessence, restricting the term to systems of time-dependent 
fields
with $w < -1/3$.  The model of the previous section does not satisfy
this criterion.  Instead, in considering the TCC and quintessence,
we focus on models with exponential
potentials yielding suitable $w$.   For this and other systems which truly exhibit quintessence, precise
statements require understanding of aspects of quantum gravity, but,
if anything, tunneling rates are likely further suppressed.

Indeed, from the work of Coleman and DeLuccia\cite{cdl}, it is known in the thin wall case that inclusion of gravitation further suppresses tunneling, and that there is no semiclassical tunneling if the radius of the would-be anti-deSitter universe is smaller than the would-be bubble size in the absence
of gravity.  In the absence of gravity, we argued for the models of the previous
section that the bubbles would be quite large, of order $\phi_0/\sqrt{V_0}$, while the tunneling amplitude is of order $e^{-\phi_0^4/V_0}$
Gravitational effects may not be within our control, but this estimate
is likely to provide some guidance as to tunneling rate.  For quintessence, in particular, the kinetic and potential terms in the action are of
comparable importance, so we might expect that neglect of the potential term would yield an ${\cal(O}(1)$ correction to the (large) exponential
factor
We need to ask:  How large is $\phi_0$?

\subsection{How large is $\phi_0$?}

The value of the field, $\phi_0$, at any given time, will control
the tunneling amplitude.   It is of interest to ask how large $\phi_0$ would be if the {\it present} universe is described by
quintessence.
Again,without good control of the quantum theory, this may be a hard
question to answer.  But the exponential potential is instructive, and strongly suggests that the tunneling amplitude is {\it extremely}
suppressed if quintessence describes the observed dark energy.
Note that in a landscape picture of quintessence, such
states would have to be quite common, so the sort of estimates
we are making here would be fairly typical.

So consider a field, $\phi$, with a nearly canonical
kinetic terms and potential
\beq
\label{exponentialpotential}
V = \Lambda^4 e^{-\lambda \phi}
\eeq
If $\Lambda = {\rm TeV}$, say, then $\lambda \phi \sim \log(10^{59})$ to
describe the current dark energy.  We know that $\lambda$ can't be too large for quintessence, probably not larger than $1 \over \sqrt{3}$ so
$\phi$ has to be something like $200$ in Planck units (note that if $\Lambda$ is larger than $TeV$ scales, as one might expect, $\phi_0$ is
larger still).  So $\phi_0^4/ V_0$ 
which is the action of the critical bubble without gravity,
is potentially huge.  E.g. if $V_0 \sim M_p^4$, then the bubble is
huge, as is the bounce action.  The would-be initial bubble itself is
two orders of magnitude larger than the AdS radius, suggestive, following
\cite{cdl} that the tunneling amplitude may vanish altogether.


We cannot make completely reliable statement for the strongly coupled system we are considering, but we would be surprised if the gravitational
result were wildly different from the result neglecting general
relativity.  Indeed, because of the large size of the critical bubble
in the semiclassical treatment, as we have said, we think it likely that the tunneling
amplitude simply vanishes for $\phi_0$ sufficiently large, and in particular for $\phi_0$ as large as required to account for the presently observed dark energy.

\section{Conclusions:  Plausibility of the Conjectures}
\label{conclusions}

Our results suggest a possible conflict between the TCC and conventional (albeit
highly speculative) conjectures about the string landscape.  If we abandon the de Sitter Swampland Conjecture, then the TCC either limits the extent in time of any superluminal
expansion, or provides a significant constraint on possible metstable de Sitter states.  While little is known about such would-be states (though there are plausible conjectures, for example \cite{kklt}, it would seem to rule out, for example, states with positive cosmological constant and even
{\it very} approximate supersymmetry\cite{dinefestucciaamorisse}.  If we hold to the conjecture, so that the observed dark energy is a form of quintessence, than  it is possible that the current 
rapid expansion might be relatively short lived, followed by a universe with $p \approx \rho$.  Alternatively, quintessence might persist, and these states
are extremely long lived once the field has evolved to a region
with very small cosmological constant. This follows if, as one might expect, there are some
modest number of negative cosmological constant states accessible to
the system.  validity of the TCC implies this quintessence state must be surrounded by some sort of dense set of AdS minima, or we have to be extremely lucky that it there is a well placed such state nearby.  

There is another possible way out.  If the parameter $\Lambda$ in
equation \ref{exponentialpotential} is {\it extremely} small, one might avoid the necessity
for large $\phi$.  This would imply a tuning condition comparable to
the usual one for the cosmological constant.  It is not clear to us whether
this might have a straightforward anthropic explanation.  We leave to the
reader the question of how plausible this might seem.

It appears most likely that either the landscape picture does not hold, or that the TCC is not valid and the theory somehow escapes the puzzles associated with the growth of subplanckian fluctuations to horizon size.

\vskip 1cm
\noindent
\noindent
{\bf Acknowledgements:}  
This work was supported in part by the U.S. Department of Energy grant number DE-FG02-04ER41286.  

\bibliography{tcc}{}
\bibliographystyle{utphys}

\end{document}